# How interband pairing increases the superconductive transition temperature

E.A. Mazur[1], V.M. Dubovik[1]

[1]National Research Nuclear University "Moscow Engineering Physics Institute", Moscow, Russia

Two-band and multi-band materials such as magnesium diboride and pnictides open new perspectives in the study of high-temperature properties of materials [1, 2]. It is believed that the high $T_c$ value in the case of the EP mechanism of superconductivity is displayed with Eliashberg strong-coupling theory [3-6] only with unreasonably high EP interaction constant $\lambda \geq 3$. In fact, at high EP coupling constant $\lambda > 2$ an another version of the theory of the EP systems [7] should be applied instead of the Migdal-Eliashberg theory. However, it was found that the actual EP interaction constant $\lambda$ in each of the bands does not exceed unity in pnictides $\lambda < 1$ (see [8-10]). In Refs. [11-15] it was shown that the reconstruction of the real Re$\Sigma$ and imaginary Im$\Sigma$ parts of the Green's function (GF) self-energy part (SP) in the case of the strong coupling is not limited to the frequency region $\omega$ approximately equal to the average phonon frequency $\omega_D$, and spreads to the larger range of frequency area $\omega \gg \omega_D$. As a result, the EP interaction modifies the GF, including its anomalous part, at a considerable distance from the Fermi surface energy in terms of the Debye phonon frequency, and not only in the vicinity of the Fermi surface $\mu - \omega_D < \omega < \mu + \omega_D$. Here with $\mu$ the chemical potential is denoted.

The aim of the present work is to study the question to what degree the experimental results for $T_c$ in two-gap materials, e.g., pnictides (see.[1,16] and references therein) are reproduced with the EP interaction and what contribution thus remains to the "area of responsibility" of the electron-electron interaction. For this purpose, in the present work a generalized version of the Migdal-Eliashberg theory for two-band materials with centers of the bands located in areas close to the same points of the reciprocal space, in particular, pnictides [16], for a nonzero temperature $T \neq 0$, is investigated in the case of two-band representation analogous to the Nambu representation for the single-band case. The developed theory describes the effects of the finite width of the electron bands and allows us to consider the effects of the variable density of electron states within the bands. The theory takes into account additionally the effect of electron-hole nonequivalence arising due to asymmetric position of the chemical potential relative to the bottom and top of the bands, as well as the nature of the two-band system. Bringing full list of works on the calculation of $T_c$ for the two-band theory seems to be too difficult, so the authors refer the reader to the existing reviews and recent works [3-5, 8-10].

Given all the above, we consider a two-band EP system with Hamiltonian $\hat{H}$ which includes the electron component $\hat{H}_e$, ion component $\hat{H}_i$, and component corresponding to electron-ion interaction in the harmonic approximation $\hat{H}_{e-i}$. Electron GF $\hat{G}$ in matrix form is defined as $\hat{G} = -<T\Psi(x)\Psi^+(x')>$ where conventional electron creation and annihilation operators are included in the generalized to the case of two-band operators Nambu form. Writing the standard equations of motion for the electron wave functions and averaging it with the Hamiltonian $\hat{H}$, we obtain the equation for the electron GF. Concentration of electrons is assumed to be small, so the screening effects of the EP interaction can be neglected due to the weak electron screening. The behavior of the matrix vertex $\hat{\Gamma}$ includes the effects of electron-electron correlation. In what follows, we do not write explicitly the first electron-electron contribution $\hat{\Sigma}_{el-el}(x,x')$, having it however in mind and considering by the behavior of the vertex $\hat{\Gamma}$ and $\hat{\Sigma}_{el-el}(x,x')$ all earlier studied (see e.g. [17]) effects of the electron-electron correlations and the effects of the interaction of electrons through spin fluctuations in two-band materials. Let us consider the interband pairing of electrons in the two-band EP system. In contrast to the single-band case the temperature electron GF in the two-band model is a $4\times4$ matrix put together by means of the creation $\psi^+_{i\alpha}(r)$ and destruction $\psi_{i\alpha}(r)$ operators of the $i$ - band electron ($i = 1, 2$) at the point $x = (r, t)$, spin projection are indicated with $\alpha$. GF for the two-band EP system can be found with the known diagram technique relation $\hat{g}^{-1} = \hat{g}_0^{-1} - \hat{\Sigma}$, where $\hat{g}_0^{-1}$ is the zero approximation inverse GF and $\hat{\Sigma}$ - matrix irreducible SP for the two-zone EP system. SP. $\hat{\Sigma}$ with neglect of the pairing of electrons in each zone separately and with neglect of all the effects of the renormalization of the chemical potential due to the interactions in each of the bands and the interband interactions can be represented as

$$\hat{\Sigma} = \begin{pmatrix} (1-Z_1)i\omega_n & 0 & 0 & \phi_{12} \\ 0 & (1-Z_1)i\omega_n & \phi_{12} & 0 \\ 0 & \phi_{12}* & (1-Z_2)i\omega_n & 0 \\ \phi_{12}* & 0 & 0 & (1-Z_2)i\omega_n \end{pmatrix} \qquad (1)$$

where $\phi_{12}$ is responsible for the pairing of two electrons of different zones. In our EP system has only one interband order parameter. Thus, we don't consider the situation with the coherent interaction of the order parameters of the two bands, first discussed in Refs. [18], [19], when in the EP system there are interfering order parameters of the first and second bands. The situation with the two interfering order parameters of two bands was considered in a number of subsequent works, for example, when applied to pnyctides [8] and magnesium diboride [1], however, the authors know of no studies in which the effects of electron pairing of two different bands have been studied. We assume to be nonzero only the GF of the different band electrons, $i \neq j$, with opposite spin moments, that is, we assume that the GF of the same band electrons with opposite spins are equal zero and the GF of the different band electrons, but with the same spins are also supposed to be equal zero. Then in the matrix $\hat{g}$ only elements located on its two diagonals will be not equal zero.

It is easy to find their explicit form by means of relation (1) and known relations for components of the matrix $\hat{g}_0^{-1}$, for example, near $T_c$ neglecting small contribution $\phi_{12}$ compared with the first term in the denominator, we have

$$g_{14} = \frac{\phi_{12}}{(Z_1 i\omega_n - \xi_{1p})(Z_2 i\omega_n + \xi_{2p})}. \qquad (2)$$

Let us write down the standard equation for the SP elements of the electron GF [20, 21], for example, $\Sigma_{14}$ in the temperature techniques. We use the spectral decomposition of the electron and phonon GF's and make the standard summation over the $\omega_n$ frequency. Take into account also the known connection of the spectral density $a(p, z)$ with the retarded GF $g(p, z)$: $a(p, z) = -2\text{Im } g(p, z)$. Make analytic continuation from the imaginary axis to the real axis, $i\omega_n \to \omega + i\delta$, and average the left and right sides of getting equation in all directions of the first band electron momentum on the energy surface $\xi_1$, whereupon $\varphi_{12}$ depends only on two variables $\xi_1$ and $\omega$. Leave in the summing up the phonon modes only one term, that corresponds to undamped modes of the phonon spectrum $b(\vec{q}, z) = 2\pi\{\delta(z - \omega_0(\vec{q})) - \delta(z + \omega_0(\vec{q}))\}$. As a result we obtain the equation for the interband order parameter $\varphi_{12}(\xi_1, \omega)$, bearing in mind that $\int_S \frac{d^2\vec{p}'}{v_{\vec{p}'}} = N(\xi)$, where $N(\xi)$ is the electron states density at the surface $\xi$ = const:

$$\varphi_{12}(\xi_1,\omega) = \int d\xi_2 \times$$

$$\times \int_{-\infty}^{\infty} \frac{dz'}{2\pi} \int_0^{\infty} dz \alpha^2(z,\xi_1,\xi_2) F(z,\xi_1,\xi_2) \left[ \frac{th\frac{z'}{2T} + cth\frac{z}{2T}}{z'+z-\omega-i\delta} - \frac{th\frac{z'}{2T} - cth\frac{z}{2T}}{z'-z-\omega-i\delta} \right] Img_{14}(\xi_1(\xi_2),\xi_2,z'), \qquad (3)$$

where the EP interaction spectral function has the form

$$\alpha^2(z,\xi_1,\xi_2) F(z,\xi_1,\xi_2) = \frac{\int_{S(\xi_1)} \frac{d^2\vec{p}}{v_{\vec{p}}} \int_{S(\xi_2)} \frac{d^2\vec{p}'}{v_{\vec{p}'}} \sum_j |g_j(\vec{p},\vec{p}')|^2 \delta(z-\omega_0(\vec{q}))}{\int_{S(\xi_1)} \frac{d^2\vec{p}}{v_{\vec{p}}} \int_{S(\xi_2)} \frac{d^2\vec{p}'}{v_{\vec{p}'}}} N_2(\xi_2) \qquad (4)$$

and $\xi_1 = \frac{p^2}{2m_1} - \mu$, $\xi_2 = \frac{p^2}{2m_2} + \Delta - \mu$, $\Delta$ is the energy shift of the lower boundaries of the two bands to each other, $g_j$ is the EP interaction matrix element, $\xi_2 = E_2 - \mu$ is the energy measured from the Fermi surface in the second band, $\int_S \frac{d^2\vec{p}'}{v_{\vec{p}'}}$ is the integral over the constant energy surface $\xi_2$ = const, which hasn't by no means coincide with the Fermi surface, and $v_\mathbf{p}$ is the electron velocity on this surface. Thus, in spite of the very general nature of the formulas obtained, we will carry out the calculations for bands centered at the same point in momentum space and shifted with the energy separation $\Delta$ one from the other. This situation, in particular, is implemented in the pnictides (Ref. [15]), where the interband EP interaction constant of the order parameters in these materials is suspected to be low [8]. Such a constant coincides in no way with the EP pairing constant for carriers of the two bands being actually used in the present work. At low frequencies $Im\varphi_{12} \ll Re\varphi_{12}$. If we assume a weak dependence $g$ on $\xi$, and hence $\varphi_{12}$ and $g_{14}$ on $\xi$, and put $Z_1 = Z_2 = 1$, then for $Re\varphi_{12}$ we obtain the following equation

$$Re\phi_{12}(\xi_1,\omega) = -\int d\xi_2 \int_{-\infty}^{\infty} \frac{dz'}{2} \int_0^{\infty} dz(\alpha^2 F) \left[ \frac{th\frac{z'}{2T} + cth\frac{z}{2T}}{z'+z-\omega} - \frac{th\frac{z'}{2T} - cth\frac{z}{2T}}{z'-z-\omega} \right] \times$$

$$\times Re\phi_{12}(\xi_1(\xi_2),\xi_2,z') \left[ \delta(z'-\xi_1(\xi_2)) \frac{P}{(z'+\xi_2)} + \delta(z'+\xi_2) \frac{P}{(z'-\xi_1(\xi_2))} \right]. \qquad (5)$$

(5) shows that both integral equations for the real and imaginary parts of the order parameter, $Re\phi_{12}(\xi_1,\omega)$ and $Im\phi_{12}(\xi_1,\omega)$, contain two integral expressions with different kernels, in distinction from the usual one-band situation [4-6], [15], where the order parameter satisfies one integral equation with a single kernel. In each relation one of the two integral expressions is responsible for the EP renormalization of the order parameter due to the interaction of the first band electron in a pair with phonons, while the second integral expression of the two corresponds to a EP renormalization of the order parameter due to the interaction with phonons of the electron from other band included in the pair. Assuming with account of what was said above the dependence of $\varphi_{12}$ on $\xi$ and $z$ to be weak, one can remove $Re\varphi_{12}$ from under the integral in the right-hand side of (5). We shall integrate with respect to $z$ using the Einstein model of the phonon spectrum ($\omega_0$ = const) and introducing the dimensionless constant of the EF interaction: $\lambda = 2\int \{\alpha^2(z,\xi_1,\xi_2(z'))F(z,\xi_1,\xi_2(z'))/z\}dz$. For this model the spectral function of the EP interaction we will write following way:
$\alpha^2(z,\xi_1,\xi_2(z'))F(z,\xi_1,\xi_2(z')) \approx \lambda\omega_0\delta(z-\omega_0)/2$. Then the equation for determining $T_c$ should be written in the form

$$1 + \frac{\lambda\omega_0}{2}\int d\xi_2 \int_{-\infty}^{\infty} dz' \left[\frac{th\frac{z'}{2T}+cth\frac{\omega_0}{2T}}{z'+\omega_0-\omega} - \frac{th\frac{z'}{2T}-cth\frac{\omega_0}{2T}}{z'-\omega_0-\omega}\right] \times \qquad (6)$$

$$\times \left[\delta(z'-\xi_1(\xi_2))\frac{P}{(z'+\xi_2)} + \delta(z'+\xi_2)\frac{P}{(z'-\xi_1(\xi_2))}\right] = 0.$$

Assuming the frequency $\omega$ to be small in comparison with $\omega_0$, we split the integral in (6) into two integrals. Since $\xi_1(\xi_2)$ looks like that: $\xi_1(\xi_2) = \frac{m_2}{m_1}\left[\xi_2 - \mu(\frac{m_1}{m_2}-1) - \Delta\right]$, then $\delta(z'-\xi_1(\xi_2)) = \frac{m_1}{m_2}\delta\left[\xi_2 - \mu(\frac{m_1}{m_2}-1) - \Delta - z'\frac{m_1}{m_2}\right]$. Integrating with respect to $\xi_2$ gives the first integral, containing in (6), in the form

$$\frac{\lambda\omega_0}{2}\frac{m_1}{(m_1+m_2)}\int_{\xi_{1\min}}^{\xi_{1\max}} dz' \left[\frac{th\frac{z'}{2T}+cth\frac{\omega_0}{2T}}{z'+\omega_0} - \frac{th\frac{z'}{2T}-cth\frac{\omega_0}{2T}}{z'-\omega_0}\right]\frac{1}{[z'+\mu\frac{(m_1-m_2)}{(m_1+m_2)}+\Delta\frac{m_2}{(m_1+m_2)}]} \qquad (7)$$

In the second integral the integration over $\xi_2$ with account of $\delta(\xi_2+z')$ given the fact, that $z'-\xi_1(-z') = \frac{(m_1+m_2)}{m_2}[z'+\mu\frac{(m_1-m_2)}{(m_1+m_2)}+\Delta\frac{m_2}{(m_1+m_2)}]$, gives the same expression (7) but with different limits in the integral over $z'$. Let us see to what are equal these limits in the first and second integrals. In the model adopted by us in the case of strong electron EP coupling the pairing in two bands does not occur near the Fermi surface and takes place throughout the depth of these bands (Fig. 1, 2). The implementation of the two-band pairing near the Fermi surface only should lead to the difficulty to satisfy the condition of zero total momentum of the Cooper pair (Fig. 2).

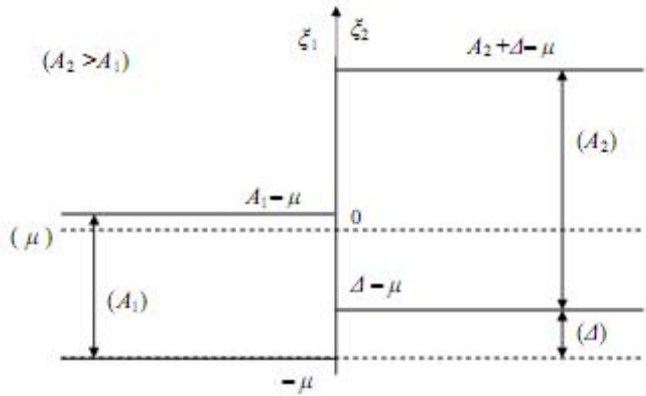

Fig.1: Diagram of two energy bands of electrons. $\xi_{1(2)}$ is the electron energy of $1^{st}$ ($2^{nd}$) - bands, measured from the chemical potential $\mu$; $A_{1(2)}$ is the width of $1^{st}$ ($2^{nd}$) - band; $\Delta$ is the distance (in energy) between the bottom of $2^{nd}$ and $1^{st}$ bands respectively.

Fig.2: Energy surfaces of electrons of $1^{st}$ and $2^{nd}$ bands in momentum space. We consider the pairing of electrons from the first band with mass $m_1$ and momentum $p$ with electrons from the second band with a mass $m_2 < m_1$ and

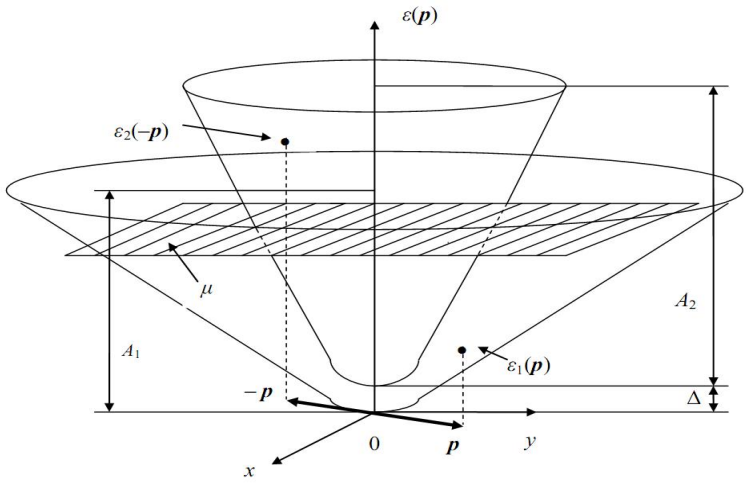

momentum - **p**. Vectors **p** and -**p** lie in the same plane $(p_x, p_y)$, and the corresponding energies $\varepsilon_1(\mathbf{p})$ and $\varepsilon_2(-\mathbf{p})$ belong to different isoenergetic surfaces.

This means that the modules of the electron momentums from the first and the second bands must be equal. In the momentum space corresponding areas are defined by the inequality: $0 \leq p \leq p_{max} = \min(p_{1rp}, p_{2rp})$ where boundary momentums are connected with the width of the bands $A_1$ and $A_2$ in following way: $p_{1rp}^2 = 2m_1 A_1$ and $p_{2rp}^2 = 2m_2 A_2$. It is easy to see that when $p_{max} = p_{1rp}$ for $\xi_1$, we have $\xi_{1max} = A_1 - \mu$. And for $\xi_2$ we have $\xi_{2max} = \frac{m_1}{m_2} A_1 + \Delta - \mu$. Similarly, when $p_{max} = p_{2rp}$ we obtain $\xi_{2max} = A_2 + \Delta - \mu$ and $\xi_{1max} = \frac{m_2}{m_1} A_2 - \mu$. Since for both bands $p_{min} = 0$ then $\xi_{1min} = -\mu$ and $\xi_{2min} = \Delta - \mu$.

Integration limits in first and second integrals are defined with conditions $z' = \xi_1$ and $z' = -\xi_2$ respectively. We introduce the following notation $\hat{A}_1, \hat{A}_2, \hat{\mu}, \hat{\Delta}$ for the values $A_1, A_2, \mu, \Delta$ expressed in the $\omega_0$ units, as well

$$x = \frac{z'}{\omega_0}, \quad k = \frac{m_2}{m_1}, \quad x_0 = \hat{\mu}\frac{1-k}{1+k} + \hat{\Delta}\frac{k}{1+k}, \quad a = \frac{\omega_0}{2T}, \quad b = \frac{\omega}{\omega_0}.$$ Let us take into account that

$$\frac{thax + ctha}{x+1} - \frac{thax - ctha}{x-1} = = -\frac{2}{x^2-1}(thax - xctha).$$ Then (7) reduces to the following equation for determining the transition temperature $T_c$ (8)

$$\frac{1+k}{\lambda} = \int_{-\hat{\mu}}^{\min\{\hat{A}_1-\hat{\mu},\,\hat{A}_2 k-\hat{\mu}\}} \frac{dx}{(x+x_0)(x^2-1)}[th(ax) - xctha] + \int_{\hat{\Delta}-\hat{\mu}}^{\min\{\hat{A}_2+\hat{\Delta}-\hat{\mu},\,\frac{\hat{A}_1}{k}+\hat{\Delta}-\hat{\mu}\}} \frac{dx}{(x-x_0)(x^2-1)}[th(ax) - xctha].$$

The values of the entered parameters will be assumed to be varying in the following intervals $0.2 < k < 3$, $2 < \hat{A}_2 < 100$, $-20 < \hat{\Delta} < 20$, $1 < \hat{\mu} < 80$. Let us take for a specific calculating the following values: $b = 0$, $\hat{\mu} \approx A_2/2$, $\hat{\Delta} \approx 0.3\hat{\mu} = 0.15 A_2$, $k \approx 0.5$, $A_2 = 10$, $A_1 = 5$. Parameter $x_0$ in this case will have the following value $x_0 = \hat{\mu}\frac{1-k}{1+k} + \hat{\Delta}\frac{k}{1+k} = \frac{A_2}{3} + 0.15 A_2 \frac{1}{3} = = \frac{A_2}{3} 1.15 \approx 0.37 A_2$. Assuming $A_1$ to be large we obtain an equation for this specific example for the $T_c$ of electron pairing for two bands centered at the same point in momentum space

$$\frac{1.5}{\lambda} = \int_{-0.5 A_2}^{0} \frac{dx}{x+x_0} \frac{1}{x^2-1}[th(ax) - xctha] + \int_{-0.35 A_2}^{0.65 A_2} \frac{dx}{x-x_0} \frac{1}{x^2-1}[th(ax) - xctha]. \quad (9)$$

Fig. 3 shows the dependence of subintegral expressions in (9) on a dimensionless variable $x$.

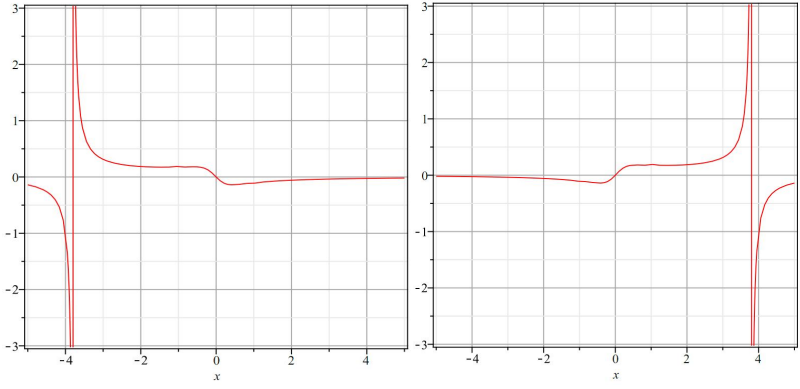

Fig. 3: The dependence on frequency *x* of the kernels of two integral expressions on the right hand side of the equation (40) to determine $T_c$ at the temperature $T = 0.15$. On Fig. 3a the kernel of the left integral contribution to (40) is presented, on Fig. 3b the right kernel of the integral contribution to (40) is presented. Frequency *x* and temperature *T* are expressed in terms of the Debye frequency $\omega_0$.

Fig. 4: The graph of interdependence of $T_c$ and interband EF coupling constant for the electrons (holes) of the two adjacent bands in terms of band parameters mentioned in the text.

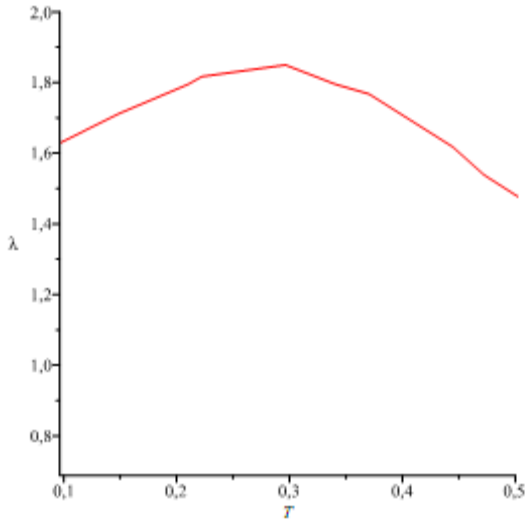

As we can see from Fig. 4, when $\lambda \approx 1.8$ two-band EP system undergoes a transition to the superconducting state at $T_c \approx 0.25\omega_0$. At $\omega_0 \approx 600K$ the transition temperature to the superconducting state reaches 150 K at $\lambda \approx 1.8$, particulaly with the set of parameter values adopted in practice: chemical potential lies near the middle of one of the bands, the energy shift of two bands centered at nearby points in momentum space equals approximately 15% of the width of the first band, and in the other twice wider band carrier effective mass is half the effective mass of the carrier in the first band. Ambiguous dependence of the superconducting transition temperature $T_c$ on the strength of the interband coupling reflects a very sophisticated effect of redistribution of EP contributions describing attraction of the two carriers belonging to two different bands with different properties, or their repulsion depending on the strength of interband interaction of the carriers, in the right side of the dual integral equation for the complex interband order parameter. It is known that the interaction of electrons through the exchange of phonons can be characterized as the attraction or repulsion depending on the frequency of the retarded electron-phonon interaction potential. The electron-phonon interaction is attractive only after a finite time interval equal to the inverse of the average phonon frequency. At the same time, the Coulomb interaction is strictly instantaneous. Usually (see., e.g. $[18-21]$) only the terms corresponding to the phonon attraction and the Coulomb repulsion are left in the equation for the energy gap. In our consideration for the energy gap both positive contributions meeting the phonon attraction as well as negative contributions corresponding phonon repulsion of two electrons are taken into account in the right-hand side of the equation (9). The resulting $T_c$ value is determined by the balance of these two contributions of opposite signs. The presence of two integrals in the right side of (9) corresponds to the attraction or repulsion in a pair of electrons due to the emission of virtual phonons by first or second electron from different bands, respectively. The structure of two kernels in the order parameter equation (9) is such that when the temperature changes not monotonic in frequency x redistribution of deposits in both kernels, corresponding both to the phonon attraction or the phonon repulsion of electrons in pairs occurs. That leads to the dependence, reflected in Fig. 4. When substituting value $\omega_0 \approx 600K$ we get somewhat overestimated $T_c \approx 150K$, that is associated with neglect in our calculations of the Coulomb pseudopotential, as well as to some overestimation of $T_c$ values obtained in the Einstein's model. The accounting of the pseudopotential of the electron-electron interaction will lead to only a very slight change in the calculated $T_c$ value. Along with the interband pairing discussed in this paper, a more general version of the Eliashberg theory should include boson pairing of carriers within each band, as well as well-known processes associated with quantum transition of pairs of carriers from one band to another [17], [18], [8], [10], [22], [23]. The resulting equations for the mass renormalization factor $Z_j(z')$ and superconducting order parameter $\varphi_j(z')$ of the $j^{-th}$ band respectively near the appearance of a nonzero interband order parameter $\varphi_{12}(z')$ will have the following form

$$\omega\left[Z_i(\omega)-1\right] = -P\sum_{j=1,2}\int_0^{+\infty} dz'\tilde{K}_{ij}^{ph}(z',\omega)N_i \frac{z'Z_i(z')}{D_i(z')}, \qquad (10)$$

$$\varphi_i(\omega) = \pi\sum_{j=1,2} P\int_{-\infty}^{+\infty} dz' N_j \cdot \tilde{K}_{ij}^{ph}(z',\omega)\frac{1}{D_j(z')}\varphi_j(z') + \pi\sum_{j=1,2} P\int_{-\infty}^{+\infty} dz' N_i \cdot \tilde{K}_{ij}^{ph}(z',\omega)\frac{1}{D_i(z')}\varphi_{12}(z'), (11)$$

$$\varphi_{12}(\omega) = \pi\int_{-\infty}^{+\infty} dz' K_{12}^{ph}(z',\omega)\Psi_{12}(z')\frac{\varphi_{12}(z')}{\sqrt{Z_1(\omega)Z_2(\omega)}} \qquad (12)$$
.

Here with $N_j$ the density of electron state in the $j^{-th}$ band is marked, $\tilde{K}_{ij}^{ph}(z',\omega)$ at $i\neq j$ corresponds to the squared interband matrix element, while the $K_{12}^{ph}(z',\omega)$ corresponds to the product of the matrix element of electron transition within the first band to the matrix element of electron transition within the second band. In the equations (10)- (13) the following notations are introduced

$$\Psi_{12}(z') = \int d\xi_2 \left[ \delta(z' - \xi_1(\xi_2)) \frac{P}{(z' + \xi_2)} + \delta(z' + \xi_2) \frac{P}{(z' - \xi_1(\xi_2))} \right], \quad D_j(z') = \left( z'^2 Z_j^2(z') + |\varphi_j(z')|^2 \right)^{\frac{1}{2}},$$

$$K_{ij}^{ph}(z', \omega) = \int_0^{+\infty} dz \alpha_{ij}^2(z) F_{ij}(z) \frac{1}{2} \left\{ \frac{th\frac{z'}{2T} + cth\frac{z}{2T}}{z' + z - \omega} - \frac{th\frac{z'}{2T} - cth\frac{z}{2T}}{z' - z - \omega} \right\}$$

Thus, the order parameter of the two-band EP system should be a quantum superposition of order parameters for each band, $\Delta_{11}$, $\Delta_{22}$, as well as the interband order parameter $\Delta_{12}$. In the compounds such as $Ba_{1-x}K_xFe_2As_2$ the band structure in the vicinity of the $\Gamma$ point of the Brillouin zone is consistent with (see, e.g. [24,25]) requirements for the zone energy shifts and for the ratio of the effective masses of carriers in these areas arising from our consideration. Unfortunately, among the existing to date «ferrous» materials it seems difficult to find a substance with high constants of the EP interaction. However, as already has been noted, the usually measured EP interaction constants of each zone individually have nothing in common with the interband EP interaction constant. Thus, purposefully selecting the materials with the band structure similar to the $Ba_{1-x}K_xFe_2As_2$ band structure and with the high interband EP interaction constant, we expect the discovery of the predicted effect. Moreover, it can be expected that with the EP interaction constant of the order unity the resulting order parameter will represent a quantum interference of the order parameter in one of the zones with only interband order parameter. This follows from the work [26], where it is shown that the interference of the order parameters of different zones for the large EP interaction constants should be suppressed. In the event $x_0 = 0$ that meets a certain balance between the chemical potential $\mu$ and the energy distance between the zones the equality $\Delta = (1 - m_1/m_2)\mu$ will be valid. With account the inequality $th(ax) \Box xctha$ that is valid for small x, (9) takes the form

$$\frac{1+k}{\lambda} = \int_{-\hat{\mu}}^{\min\{\hat{A}_1 - \hat{\mu}, \hat{A}_2 k - \hat{\mu}\}} dx \frac{th(ax)}{x} \frac{1}{(1-x^2)} + \int_{\hat{\Delta} - \hat{\mu}}^{\min\{\hat{A}_2 + \hat{\Delta} - \hat{\mu}, \frac{\hat{A}_1}{k} + \hat{\Delta} - \hat{\mu}\}} dx \frac{th(ax)}{x} \frac{1}{(1-x^2)}. \tag{13}$$

The integrand in (13) is rapidly reduced as a consequence of the $\frac{th(ax)}{x}$ behavior in this case. We neglect $x^2$ to 1 in $1 - x^2$ in (13) and restrict the integration over x with unit, that is, with the Debye frequency value, as is usually done. As a result we obtain the equation for $T_c$ in the next fairly standard form $\frac{1+k}{2\lambda} = K\left(\frac{\omega_D}{2T_c}\right)$, where $\int_0^y dx \frac{thx}{x}$ with $K(y)$ is indicated. The solution of such an equation is well known $T_c = \frac{2\gamma}{\pi} \omega_D \exp\left(-\frac{(1+k)}{2\lambda}\right)$ and describes the situation with the conventional superconductivity in the case of the weak coupling, but with the interband EP pairing. Thus, the other opportunity to detect the effect is the careful $Ba_{1-x}K_xFe_2As_2$ doping in order to precisely comply with the condition $\Delta = (1 - m_1/m_2)\mu$ under which the interband superconductivity arises even with the low EP interband coupling.

The conclusion for the existence of another family of materials with high superconducting transition temperature $T_c$ not yielding $T_c$ in cuprates from this work is emerging.